\documentclass[12pt]{iopart}

\usepackage{iopams}  

\usepackage{cite}
\usepackage{graphicx}
\usepackage{multirow}
\usepackage{bm}

\usepackage{color}

\begin{document}

\title[Extended duality relations between birth-death processes and PDE]{Extended duality relations between birth-death processes and partial differential equations}


\author{Jun Ohkubo}

\address{
Graduate School of Informatics, Kyoto University,\\
36-1, Yoshida Hon-machi, Sakyo-ku, Kyoto-shi, Kyoto 606-8501, Japan
}
\ead{ohkubo@i.kyoto-u.ac.jp}
\begin{abstract}
Duality relations between continuous-state and discrete-state stochastic processes with continuous time
have already been studied and used in various research fields.
We propose extended duality relations, which enable us
to derive discrete-state stochastic processes from arbitrary diffusion-type partial differential equations.
The derivation is based on the Doi-Peliti formalism,
and it will be clarified that additional states for the discrete-state stochastic processes
must be considered in some cases.
\end{abstract}

\maketitle

\section{Introduction}
\label{sec_introduction}

Duality is a widely used technique to investigate interacting particle systems
(for example, see \cite{Liggett_book}).
Using the duality concept,
a stochastic process may be connected to a different kind of stochastic processes;
for example, duality relations between 
a continuous-state stochastic process (a stochastic differential equation) 
and
a discrete-state stochastic process (a birth-death process)
have been much investigated in various research contexts ranging from 
population genetics \cite{Shiga1986,Mohle1999} 
to nonequilibrium heat-conduction problems \cite{Giardina2007}.

Although the construction of a dual process has mainly been performed heuristically,
more general schemes for deriving the dual process have recently been investigated \cite{Giardina2009,Ohkubo2010}.
In \cite{Ohkubo2010}, the Doi-Peliti formalism \cite{Doi1976,Doi1976a,Peliti1985}
has been used to derive the duality relations
between continuous-state and discrete-state stochastic processes.
As a consequence, it is possible to derive partial differential equations
from arbitrary birth-death processes describing chemical reactions or ecological systems.
Note that for a second-order partial differential equation
it is sometimes possible to derive a corresponding stochastic differential equation,
and hence we can derive the duality relations
between the continuous-state and discrete-state stochastic processes.
However, the reverse direction has not been clear.
That is, it was unclear whether birth-death processes 
can be derived from `arbitrary' partial differential equations or not;
the Doi-Peliti formalism can derive a kind of linear operator from partial differential equations,
but there is no guarantee that the derived operator can be interpreted
as being that of birth-death processes.

In the present paper, we propose extensions of the duality concept,
which enable us to derive a birth-death process
from a diffusion-type partial differential equation.
In order to connect the partial differential equation
with a discrete-state `stochastic' process,
additional states should in some cases be introduced.
Using the extended duality relations,
it is possible to evaluate moments of the continuous-state stochastic processes
or partial differential equations
from the solutions of the discrete-state stochastic processes.


The structure of the present paper is as follows.
In section 2, the conventional duality is briefly reviewed,
in which the derivation of the dual process via the Doi-Peliti formalism
is shortly explained.
Section~3 gives an extension of the duality; a weight term, so called Feynman-Kac term, is introduced.
The main claim of the present paper is discussed in section~4,
in which not only the Feynman-Kac term, but also additional states are introduced
in order to define the extended duality relations.
Section~5 gives discussions.
Although all discussions in the present paper will be performed by using some specific examples
for clarity,
the discussions can be applied to various cases;
as for the applicability of the scheme, see section 5.

\section{Conventional duality}

In this section, we briefly explain the conventional duality relations
between continuous-state and discrete-state stochastic processes.
The discussions and derivations are basically based on \cite{Ohkubo2010}.
In \cite{Ohkubo2010}, it was pointed out that
the correspondence between the generating function approach
and the Doi-Peliti formalism is available 
to derive the duality.
However, we noticed that there is no need to use the correspondence explicitly.
here, We give the derivation without using the generating function approach.

\subsection{Definition of the duality}
\label{sec_definition_of_duality}

The conventional duality between a continuous-state stochastic process
and a birth-death process is defined as follows.
Suppose that $(z_t)_{t \geq 0} \in \mathbb{R}$ is 
a continuous-state and continuous-time Markov process,
and that $(n_t)_{t \geq 0} \in \mathbb{N}$ is a birth-death process.
Let $\mathbb{E}_{n}$ denote the expectation
with respect to the process $(n_t)_{t \geq 0}$ which starts from $n_0$.
The process $(n_t)_{t \geq 0}$ is said to be \textit{dual} to $(z_t)_{t \geq 0}$
with respect to a duality function 
$D: \mathbb{R} \times \mathbb{N} \to \mathbb{R}$
if for all $z \in \mathbb{R}$, $n \in \mathbb{N}$ and $t \geq 0$ we have
\begin{eqnarray}
\mathbb{E}_{n} \left[ D(z_0,n_t)  \right] = \mathbb{E}_{z} \left[ D(z_t,n_0) \right],
\label{eq_duality_original}
\end{eqnarray}
where $\mathbb{E}_{z}$ is the expectation
in the process $(z_t)_{t \geq 0}$ starting from $z_0$.

\subsection{Problem settings}

In order to see the duality relations in more explicit forms,
we consider here a concrete partial differential equation,
which is related to the stochastic Fisher and
Kolmogorov-Petrovsky-Piscounov (sFKPP) equation;
the sFKPP equation plays an important role in the study of the front-propagation problems
 \cite{Brunet1997,Pechenik1999,Panja2004,Brunet2006,Doering2003},
and even in a QCD context \cite{Munier2006}.
The partial differential equation for $0 \le z \le 1$ with constant parameters $\gamma$ and $\sigma$,
\begin{eqnarray}
\frac{\partial}{\partial t} p(z,t) 
= - \frac{\partial}{\partial z} \left[ -\gamma z (1-z) p(z,t) \right] 
+ \frac{1}{2} \frac{\partial^2}{\partial z^2} \left[ \sigma^2 z (1-z) p(z,t) \right],
\label{eq_pde_1}
\end{eqnarray}
can be interpreted as the following stochastic differential equation:
\begin{eqnarray}
\rmd z = - \gamma z(1-z) \rmd t + \sigma \sqrt{z(1-z)} \, \rmd W(t),
\label{eq_sde_1}
\end{eqnarray}
where $W(t)$ is a Wiener process and we used the Ito-type stochastic differential equation.
Note that when a variable transformation $u(t) \equiv 1 - z(t)$ is employed,
the original sFKPP equation is recovered \cite{Ohkubo2010}.
It has been shown that the $n$-th moment of the stochastic differential equation in \eref{eq_sde_1}
or the partial differential equation in \eref{eq_pde_1},
defined as
\begin{eqnarray}
\mathrm{E}_{z}[z_t^n] \equiv \int_{-\infty}^\infty \rmd z \, p(z,t) z^n,
\end{eqnarray}
can be evaluated without solving these equations directly.
That is, the evaluation is performed from a corresponding birth-death process
via the duality relations introduced in \eref{eq_duality_original}.

\subsection{Derivation of the dual process and duality function}
\label{sec_dual_process_1}

The problem here is how to fine the corresponding dual process and the duality function.
In order to find them in a unified way,
we first introduce creation and annihilation operators 
in order to discuss stochastic processes \cite{Doi1976,Doi1976a,Peliti1985}.
This method is called the Doi-Peliti formalism, the second quantization method,
or the field-theoretic method, and so on,
and it has been used to investigate stochastic processes
mainly from the view point of perturbation calculations or renormalization group methods
(for example, see \cite{Tauber2005}).

In the Doi-Peliti formalism, the following bosonic commutation relation is introduced:
\begin{eqnarray}
[a,a^\dagger] \equiv a a^\dagger - a^\dagger a = 1, \quad [a, a] = [a^\dagger,a^\dagger] = 0,
\end{eqnarray}
where $a^\dagger$ is the creation operator,
and $a$ the annihilation operator.
Each operator works on a ket vector in Fock space $| n \rangle$ as follows:
\begin{eqnarray}
a^\dagger | n \rangle = | n+1 \rangle, \quad a | n \rangle = n | n-1 \rangle,
\end{eqnarray}
and the vacuum state $|0\rangle$ is characterized by $a | 0 \rangle = 0$.
The bra vector $\langle m |$ is naturally introduced
from the inner product defined as
\begin{eqnarray}
\langle m | n \rangle = \delta_{m,n} n!,
\label{eq_DP_inner_product}
\end{eqnarray}
where $\delta_{m,n}$ is the Kronecker delta.

In order to connect the partial differential equation in \eref{eq_pde_1}
with the discussions relating to the Doi-Peliti formalism,
it is convenient to introduce the following linear operators:
\begin{eqnarray}
L\left( z,\frac{\partial}{\partial z} \right) 
= - \gamma z (1-z) \frac{\partial}{\partial z} + \frac{\sigma^2}{2}z(1-z) \frac{\partial^2}{\partial z^2}
\end{eqnarray}
and
\begin{eqnarray}
L^{*} \left( z, \frac{\partial}{\partial z} \right)
= - \frac{\partial}{\partial z} \left[ -\gamma z (1-z) \right]
+ \frac{1}{2} \frac{\partial^2}{\partial z^2} \left[ \sigma^2 z(1-z) \right],
\end{eqnarray}
i.e., $L^{*}$ is the adjoint operator of $L$.
Hence, the partial differential equation in \eref{eq_pde_1}
can be rewritten as follows:
\begin{eqnarray}
\frac{\partial}{\partial t} p(z,t) = 
L^{*}\left(z,\frac{\partial}{\partial z} \right) p(z,t).
\end{eqnarray}
Next, the following state vector $\langle p(t) |$ 
characterized by the probability density $p(z,t)$ is introduced:
\begin{eqnarray}
\langle p(t) | \equiv \int_{-\infty}^\infty \rmd z \, p(z,t) \langle z |,
\end{eqnarray}
where $\langle z |$ is a coherent state of $a^\dagger$, i.e., 
\begin{eqnarray}
\langle z | \equiv \langle 0 | \mathrm{e}^{z a}, 
\end{eqnarray}
which satisfies 
\begin{eqnarray}
\langle z | a^\dagger = z \langle z |,
\end{eqnarray}
and $z$ is assumed to be a real variable.

We here construct the linear operator
$L\left( a^\dagger, a \right)$
by simply replacing $z$ and $\frac{\partial}{\partial z}$
with $a^\dagger$ and $a$, respectively;
\begin{eqnarray}
L(a^\dagger, a) = - \gamma a^\dagger (1-a^\dagger) a + \frac{\sigma^2}{2} a^\dagger (1-a^\dagger) a^2.
\end{eqnarray}
Note that the linear operator $L(a^\dagger,a)$ is written in the normal order;
all annihilation operators $a$ are to the right of all creation operators in the products.
Then, using the following formula,
\begin{eqnarray}
\langle z | (a^\dagger)^k a^l 
= \langle z | z^k a^l 
= z^k \langle 0 | \left( \frac{\partial}{\partial z} \right)^l \rme^{za} 
= z^k \left( \frac{\partial}{\partial z} \right)^l \langle z |,
\end{eqnarray}
we have the following identity:
\begin{eqnarray}
\langle p(t) | L\left( a^\dagger, a \right)
&= \int_{-\infty}^{\infty} \rmd z \, p(z,t) \langle z | L\left( a^\dagger, a \right) \nonumber \\
&= \int_{-\infty}^{\infty} \rmd z \, p(z,t) 
L\left( z, \frac{\partial}{\partial z}\right)
\langle z |  \nonumber \\
&= \int_{-\infty}^{\infty} \rmd z \, 
\left[ L^*\left( z, \frac{\partial}{\partial z}\right) p(z,t) \right] 
\langle z |.
\end{eqnarray}
Hence, the solution of the partial differential equation in \eref{eq_pde_1}
is obtained formally from the following state vector:
\begin{eqnarray}
\langle p(t) | 
= \langle p(0) | \exp \left[ L\left( a^\dagger, a \right) t \right].
\end{eqnarray}
Note that it is also possible to consider cases with time-dependent coefficients.
For simplicity, we here restrict ourselves to the case with time-independent coefficients.

Next, a state vector spanned by the basis of the Doi-Peliti formalism
is introduced as follows:
\begin{eqnarray}
| P(t) \rangle \equiv \sum_{n=0}^\infty P(n,t) | n \rangle,
\end{eqnarray}
where we assume that $P(n,t)$ is adequately normalised as
\begin{eqnarray}
\sum_{n=0}^\infty P(n,t) = 1.
\end{eqnarray}
Assuming that the action of the linear operator $L(a^\dagger, a)$ to the state vector $| P(t) \rangle$
gives the time-evolution of the state vector,
i.e.,
\begin{eqnarray}
\exp\left[ L(a^\dagger,a) t\right] | P(0) \rangle = | P(t) \rangle,
\label{eq_should_be_verified}
\end{eqnarray}
the following duality is obtained:
\begin{eqnarray}
\langle p(0) | P(t) \rangle
= \langle p(0) | \exp[L(a^\dagger,a) t] | P(0) \rangle
= \langle p(t) | P(0) \rangle.
\label{eq_duality_pre_1}
\end{eqnarray}
Hence, the duality in \eref{eq_duality_pre_1}
is explicitly written as
\begin{eqnarray}
\int_{-\infty}^\infty \rmd z \sum_{n=0}^\infty p(z,0) P(n,t) z^n
= \int_{-\infty}^\infty \rmd z \sum_{n=0}^\infty p(z,t) P(n,0) z^n,
\end{eqnarray}
where we used the following identity:
\begin{eqnarray}
\fl
\langle z | n \rangle 
= \langle 0 | \sum_{m=0}^\infty \frac{1}{m!} z^m a^m | n \rangle
= \sum_{m=0}^\infty \frac{1}{m!} z^m \langle m | n \rangle
= \sum_{m=0}^\infty \frac{1}{m!} z^m \delta_{m,n} m!
= z^n.
\end{eqnarray}
We therefore conclude that
the duality function is given as
\begin{eqnarray}
D(z,n) = z^n.
\end{eqnarray}
If we set the initial conditions for $n$ and $z$ as a Kronecker delta function
and a Dirac delta function respectively,
the duality relation in \eref{eq_duality_original} is recovered;
using the explicit expression for the duality function,
we have the following duality relation:
\begin{eqnarray}
\mathbb{E}_{n} \left[ z_0^{n_t} \right]
= \mathbb{E}_{z} \left[ z_t^{n_0} \right].
\label{eq_duality_1}
\end{eqnarray}

The remaining problem in the above discussions is the verification
of \eref{eq_should_be_verified}.
Is it possible to interpret \eref{eq_should_be_verified} as the time evolution
of the probability distribution $P(n,t)$?
If not, \eref{eq_should_be_verified} is not valid and 
the duality relation is not derived.

In order to check the validity, 
we split the time-evolution operator $\exp(L(a^\dagger,a) t)$
as a product of $\exp(L(a^\dagger,a) \Delta t)$ with small $\Delta t$.
The expansion of $\exp(L(a^\dagger,a) \Delta t)$ gives
\begin{eqnarray}
\fl
\exp( L(a^\dagger,a) \Delta t) 
&\simeq 1 + L(a^\dagger,a) \Delta t \nonumber \\
&= \left[ 1 - \gamma a^\dagger a \Delta t - \frac{\sigma^2}{2} (a^\dagger)^2 a^2 \Delta t \right]
+ \gamma (a^\dagger)^2 a \Delta t + \frac{\sigma^2}{2} a^\dagger a^2 \Delta t,
\label{eq_expansion_of_L_1}
\end{eqnarray}
where $1$ is the identity operator.
Noting $a^\dagger a | n \rangle = n | n \rangle$,
the expansion in \eref{eq_expansion_of_L_1} can be interpreted as follows.
Assume that the current state is $| n \rangle$.
The second term $\gamma (a^\dagger)^2 a \Delta t$ 
means that there is a state change $n \to n+1$ with probability $\gamma n \Delta t$.
The third term $(\sigma^2/2)a^\dagger a^2 \Delta t$ corresponds
to a change $n \to n-1$ with probability $(\sigma^2/2) n (n-1) \Delta t$.
The first term corresponds to the case in which no transition occurs.
The sum of the probabilities for these events is equal to one,
and hence the operator $\exp(L(a^\dagger,a) \Delta t)$ 
gives the time evolution for the probability distribution.
More explicitly, we consider the following time-evolution equation
for the state vector $| P(t) \rangle$:
\begin{eqnarray}
\frac{\partial}{\partial t} | P(t) \rangle = L(a^\dagger,a) | P(t) \rangle,
\label{eq_time_evolution_for_ket_state_1}
\end{eqnarray}
which gives the following master equation for $P(n,t)$ by comparing the coefficients of
$|n\rangle$ of the left- and right-hand sides in \eref{eq_time_evolution_for_ket_state_1}:
\begin{eqnarray}
\frac{\partial}{\partial t} P(n,t) 
=&
\gamma (n-1) P(n-1,t) - \gamma n P(n,t) \nonumber \\
&+ \sigma^2 \frac{(n+1)n}{2} P(n+1,t) - \sigma^2 \frac{n(n-1)}{2}P(n,t).
\label{eq_derived_master_eq}
\end{eqnarray}
The master equation in \eref{eq_derived_master_eq}
suggests the following birth-coagulation process for particles $A$:
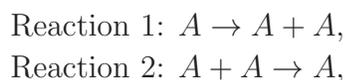
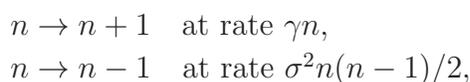
\begin{eqnarray*}
\begin{array}{l}
\textrm{Reaction 1: } A \to A+A, \\
\textrm{Reaction 2: } A + A \to A,
\end{array}
\end{eqnarray*}
i.e., 
\begin{eqnarray*}
\begin{array}{l}
n \to n+1 \quad \textrm{at rate $\gamma n$},  \\
n \to n-1 \quad \textrm{at rate $\sigma^2 n(n-1)/2$},
\end{array}
\end{eqnarray*}
where $n$ is the number of particles $A$.
Hence, we can verify \eref{eq_should_be_verified},
and the partial differential equation in \eref{eq_pde_1}
or the continuous stochastic process in \eref{eq_sde_1}
is connected to the birth-death process in \eref{eq_derived_master_eq}
via the duality relation in \eref{eq_duality_1}.

In general, the discrete-state stochastic processes are easily solved
compared with the continuous-state stochastic processes,
and actually the duality relations have been mainly used
to evaluate various quantities for the continuous-state stochastic processes (e.g., see \cite{Giardina2007}).

\section{Duality with Feynman-Kac terms}

\subsection{Problem settings}

Instead of the original problem in \eref{eq_pde_1},
we next consider the following partial differential equation:
\begin{eqnarray}
\frac{\partial}{\partial t} p(z,t) 
= - \frac{\partial}{\partial z} \left[ -\gamma z (1-z) p(z,t) \right] 
+ \frac{1}{2} \frac{\partial^2}{\partial z^2} \left[  \sigma^2 z  p(z,t) \right] .
\label{eq_pde_2}
\end{eqnarray}
Note that the coefficient in the second term in the right-hand side is different
from \eref{eq_pde_1}.
The partial differential equation can be interpreted as 
the following stochastic differential equation:
\begin{eqnarray}
\rmd z = - \gamma z(1-z) \rmd t + \sigma \sqrt{z} \, \rmd W(t).
\end{eqnarray}
The corresponding adjoint operator $L^*\left( z,\frac{\partial}{\partial z} \right)$ is
\begin{eqnarray}
L^{*}\left( z,\frac{\partial}{\partial z} \right)
= - \frac{\partial}{\partial z} \left[ -\gamma z (1-z) \right]
+ \frac{1}{2} \frac{\partial^2}{\partial z^2} \sigma^2 z,
\end{eqnarray}
and the linear operator in terms of the creation and annihilation operators is
\begin{eqnarray}
L(a^\dagger,a) = - \gamma a^\dagger (1-a^\dagger) a + \frac{\sigma^2}{2} a^\dagger a^2.
\label{eq_generator_2}
\end{eqnarray}

One may expect that the linear operator $L(a^\dagger,a)$
describes time evolutions of a certain type of stochastic process.
However, the same discussions in section~\ref{sec_dual_process_1}
do not give a stochastic process;
the probability conservation is violated
and \eref{eq_should_be_verified} cannot be verified.
In order to derive a dual stochastic process,
it is necessary to extend the definition of the duality using a weight term,
the so-called Feynman-Kac term \cite{Liggett_book}.
Actually, such extension has already been introduced \cite{Liggett_book},
and we here give a derivation of the Feynman-Kac term
in terms of the Doi-Peliti formalism.

\subsection{Duality}
\label{sec_dual_process_2}

The second term in \eref{eq_generator_2} changes the states;
it contains $a^\dagger a^2$, and then there is a state change $n \to n-1$.
However, there is no corresponding term for the probability conservation.
Hence, a simple solution for this problem is to add an additional term,
which does not change the state and complements the lack of the probability conservation.
That is, we rewrite the linear operator $L(a^\dagger,a)$ in \eref{eq_generator_2} as
\begin{eqnarray}
L(a^\dagger,a) &= - \gamma a^\dagger (1-a^\dagger) a + \frac{\sigma^2}{2} a^\dagger (1-a^\dagger) a^2
+ \frac{\sigma^2}{2} (a^\dagger)^2 a^2 \nonumber \\
&= L'(a^\dagger,a) + V(a^\dagger a),
\end{eqnarray}
where 
\begin{eqnarray}
L'(a^\dagger,a) &= - \gamma a^\dagger (1-a^\dagger) a + \frac{\sigma^2}{2} a^\dagger (1-a^\dagger) a^2
\end{eqnarray}
and
\begin{eqnarray}
V(a^\dagger a) &= \frac{\sigma^2}{2} \left[ (a^\dagger a)^2 - a^\dagger a \right].
\end{eqnarray}
The term $V(a^\dagger a)$ corresponds to the Feynman-Kac term.
Note that the operator $V(a^\dagger a)$ is expressed only in terms of $a^\dagger a$,
and therefore the operator $V(a^\dagger a)$ does not change the state $| n \rangle$.
Considering a small time interval $\Delta t$, we have
\begin{eqnarray}
\exp \left[ L(a^\dagger,a) \Delta t \right] | n \rangle
&= \exp \left[ V(a^\dagger a) \Delta t +  L'(a^\dagger,a) \Delta t \right] | n \rangle \nonumber \\
&\simeq \left( 1 + V(a^\dagger a) \Delta t +  L'(a^\dagger,a) \Delta t \right) | n \rangle \nonumber \\
&= \left( 1 + V( n) \Delta t +  L'(a^\dagger,a) \Delta t \right) | n \rangle,
\label{eq_operator_split_2}
\end{eqnarray}
and we can replace $\exp[ V(a^\dagger a) \Delta t + L'(a^\dagger,a) \Delta t ]$ with
$\exp[ V( n) \Delta t + L'(a^\dagger,a) \Delta t ]$ 
for the small time interval $\Delta t$,
which makes $\exp(V(n) \Delta t)$ a scalar value.
Hence, only the factor $\exp[L'(a^\dagger,a) \Delta t]$ 
plays a role as a time-evolution operator.
The factor $\exp[L'(a^\dagger,a) \Delta t]$  can be interpreted
as the time-evolution operator for the birth-death process in section~\ref{sec_dual_process_1},
and we have
\begin{eqnarray}
\exp\left[ L'(a^\dagger,a) t\right] | P(0) \rangle = | P(t) \rangle,
\end{eqnarray}
instead of \eref{eq_should_be_verified}.
As a consequence, the duality relation should be modified a little,
and finally the following duality relation is obtained instead of \eref{eq_duality_1}:
\begin{eqnarray}
\mathbb{E}_{n} \left[  \exp\left\{ \int_0^t V(n_{t'}) \, \rmd t' \right\} z_0^{n_t} \right]
= \mathbb{E}_{z} \left[ z_t^{n_0} \right].
\label{eq_duality_2}
\end{eqnarray}
Note that the expectation on the left hand side of \eref{eq_duality_2}
should be taken for all the paths
because the exponential of the Feynman-Kac term, $\exp(\int_0^t V(n_{t'}) \rmd t')$, depends on the path.

\section{Duality with Feynman-Kac terms and additional states}

\subsection{Problem settings}

In this section, we deal with more complicated cases,
which is seen in the following example:
\begin{eqnarray}
\frac{\partial}{\partial t} p(z,t) 
= - \frac{\partial}{\partial z} \left[ \gamma z (1-z)  p(z,t) \right]
+ \frac{1}{2} \frac{\mathrm{\partial}^2}{\mathrm{\partial} z^2} \left[ \sigma^2 z (1-z) 
p(z,t) \right],
\label{eq_pde_3}
\end{eqnarray}
which can be interpreted as
\begin{eqnarray}
\rmd z = \gamma z(1-z) \rmd t + \sigma \sqrt{z (1-z)} \, \rmd W(t).
\label{eq_sde_3}
\end{eqnarray}
The adjoint operator $L^*\left( z,\frac{\partial}{\partial z} \right)$ is written as
\begin{eqnarray}
L^{*} \left( z,\frac{\partial}{\partial z} \right)
= - \frac{\partial}{\partial z} \left[ \gamma z (1-z) \right]
+ \frac{1}{2} \frac{\partial^2}{\partial z^2} \sigma^2 \left[ (1-z)z \right],
\end{eqnarray}
and in terms of the creation and annihilation operators, $L(a^\dagger,a)$ becomes
\begin{eqnarray}
L(a^\dagger,a) &= \gamma a^\dagger (1- a^\dagger) a + \frac{\sigma^2}{2} a^\dagger (1-a^\dagger)  a^2 \nonumber \\
&= \gamma a^\dagger a - \gamma (a^\dagger)^2 a + \frac{\sigma^2}{2} a^\dagger (1-a^\dagger)  a^2.
\label{eq_generator_3}
\end{eqnarray}

Notice that there is the term, $- \gamma (a^\dagger)^2 a$, in \eref{eq_generator_3}.
This term changes the state as $n \to n+1$,
but the transition probability is negative. 
(In order to see it, use the expansion similar to \eref{eq_expansion_of_L_1}.)
It is impossible to consider such negative transition probabilities
for usual stochastic processes.
In addition, we cannot use the similar techniques introduced in section~\ref{sec_dual_process_2};
there is no way to cancel the effects of the transition with the negative probability
or to split the operators in a simple form as \eref{eq_operator_split_2}.

In the following discussions,
we introduce additional operators and an additional state.
This additional state enables us 
to adequately deal with the transitions with the negative probability.

\subsection{Additional states and duality}

In order to discuss the negative transition probability,
it is useful to introduce the following new state vectors, 
$|+\rangle$, $|-\rangle$, $\langle + |$, and $\langle -|$.
They satisfy
\begin{eqnarray}
\langle + | + \rangle = \langle - | - \rangle = 1,
\quad \langle + | - \rangle = \langle - | + \rangle = 0.
\end{eqnarray}
Next, a new operator $b$ is defined as
\begin{eqnarray}
b | + \rangle = | - \rangle, \quad b | - \rangle = | + \rangle, 
\quad \langle + | b = \langle -|, \quad \langle - | b = \langle +|.
\end{eqnarray}
We here give a mathematical comment on the above construction.
The above construction is based on the algebraic probability theory;
we can confirm that the above construction gives
an adequate algebraic probability space.
As for the details of the algebraic probability theory,
for example, see \cite{Hora_Obata_book, Ohkubo2012}.

Using the above additional states $|+\rangle$ and $|-\rangle$,
we consider the following combined stochastic process.

Firstly, a new state vector $|\psi(t)\rangle$ for the birth-death process is introduced as 
\begin{eqnarray}
| \psi(t) \rangle 
= \sum_{n=0}^\infty \sum_{s=\{+,-\}}  P(n,s,t)  \, | n , s \rangle,
\end{eqnarray}
where $P(n,s,t)$ is the probability with which
the number of particles in the birth-death process is $n$
and the sign of the state is $s \in \{+,-\}$ at time $t$,
and $|n,s\rangle = |n\rangle \otimes | s \rangle$.
Note that the operators $a^\dagger$ and $a$ act only on the particle numbers $n$,
and the operator $b$ affects only on the sign of the state $s \in \{+,-\}$.
That is,
\begin{eqnarray}
a^\dagger | n, s \rangle = | n+1, s \rangle, 
\quad a | n, s \rangle = n | n-1, s \rangle, 
\end{eqnarray}
for $s \in \{+,- \}$, and 
\begin{eqnarray}
b | n, + \rangle = | n, - \rangle,
\quad b | n, - \rangle = | n, + \rangle.
\end{eqnarray}
In addition, we define $\langle p(t)|$ as follows:
\begin{eqnarray}
\langle p(t) | = \int_{-\infty}^\infty \rmd z \, p(z,t) 
\left[ \langle z,+| - \langle z,-| \right].
\end{eqnarray}
Using the above definition,
there is no need to define the sign of the states, $+$ or $-$,
for the partial differential equations, as we will see soon.

Secondly, we construct 
a linear operator acting on the new state vectors $| \psi(t) \rangle$
based on the linear operator $L(a^\dagger,a)$ in \eref{eq_generator_3}, 
as follows:
\begin{eqnarray}
L(a^\dagger,a,b) 
& = \gamma a^\dagger a + b \gamma (a^\dagger)^2 a + \frac{\sigma^2}{2} a^\dagger (1-a^\dagger)  a^2
\nonumber \\
& = 2 \gamma a^\dagger a - \gamma a^\dagger (1 - b a^\dagger) a + \frac{\sigma^2}{2} a^\dagger (1-a^\dagger)  a^2
\nonumber \\
&= L'(a^\dagger,a,b) + V(a^\dagger a),
\label{eq_generator_new_3}
\end{eqnarray}
where 
\begin{eqnarray}
L'(a^\dagger,a,b) &= - \gamma a^\dagger (1- b a^\dagger) a + \frac{\sigma^2}{2} a^\dagger (1-a^\dagger) a^2
\end{eqnarray}
and
\begin{eqnarray}
V(a^\dagger a) &= 2 \gamma a^\dagger a.
\end{eqnarray}
Note that for the negative sign the second term in \eref{eq_generator_3} 
is replaced with the new operator $b$.

Thirdly, we check the action of $L(a^\dagger,a,b)$ for $\langle p(t)|$;
\begin{eqnarray}
\fl
\langle p(t) | L(a^\dagger,a,b) \nonumber \\
\fl 
= \int_{-\infty}^\infty \rmd z \, p(z,t) 
\left[ \langle z,+| - \langle z,-| \right] L(a^\dagger,a,b) \nonumber \\
\fl
= \int_{-\infty}^\infty \rmd z \, 
\left\{ 
- \frac{\partial}{\partial z} \left[ \gamma z p(z,t) \right] \langle z,+|
+ \frac{\partial}{\partial z} \left[ \gamma z^2 p(z,t) \right] \langle z,+| 
+ \frac{\partial}{\partial z} \left[ \gamma z p(z,t) \right] \langle z,-|
\right. \nonumber \\
\fl \qquad \qquad \quad
\left. - \frac{\partial}{\partial z} \left[ \gamma z^2 p(z,t) \right] \langle z,-| 
+ \frac{1}{2} \frac{\partial^2}{\partial z^2} 
\left[ \sigma^2 (1-z)z p(z,t)\right] \left[ \langle z,+| - \langle z,-| \right] 
\right\}.
\label{eq_for_pde_3_1}
\end{eqnarray}
On the other hands, 
\begin{eqnarray}
\frac{\rmd}{\rmd t} \langle p(t) |
= \int_{-\infty}^\infty \rmd z \, 
\left( \frac{\rmd}{\rmd t} p(z,t) \right) \left[ \langle z,+| - \langle z,-| \right].
\label{eq_for_pde_3_2}
\end{eqnarray}
Comparing the coefficients of $\langle z,+|$ (or $\langle z,-|$)
in \eref{eq_for_pde_3_1} and \eref{eq_for_pde_3_2},
we have the original partial differential equation in \eref{eq_pde_3}.
Hence, the new time-evolution operator defined in \eref{eq_generator_new_3}
adequately recovers the original problem.

Fourthly,
we investigate the action of the time-evolution operator $L'(a^\dagger,a,b)$
in \eref{eq_generator_new_3}.
(The effects of $V(a^\dagger a)$ are the same as in section~\ref{sec_dual_process_2},
i.e., a Feynman-Kac term.)
After some calculations, the following master equations for $P(n,s,t)$ are obtained:
\begin{eqnarray}
\fl
\frac{\partial}{\partial t} P(n,+,t) 
=&
\gamma (n-1) P(n-1,-,t) - \gamma n P(n,+,t) \nonumber \\
&+ \sigma^2 \frac{(n+1)n}{2} P(n+1,+,t) - \sigma^2 \frac{n(n-1)}{2}P(n,+,t), 
\label{eq_derived_master_eq_3_1} \\
\fl
\frac{\partial}{\partial t} P(n,-,t) 
=&
\gamma (n-1) P(n-1,+,t) - \gamma n P(n,-,t) \nonumber \\
&+ \sigma^2 \frac{(n+1)n}{2} P(n+1,-,t) - \sigma^2 \frac{n(n-1)}{2}P(n,-,t).
\label{eq_derived_master_eq_3_2}
\end{eqnarray}
The master equations in \eref{eq_derived_master_eq_3_1} and \eref{eq_derived_master_eq_3_2}
can be interpreted as interacting particle systems with internal states,
i.e., the sign of the states;
the birth-death process described by 
\eref{eq_derived_master_eq_3_1} and \eref{eq_derived_master_eq_3_2}
becomes as follows:
\begin{eqnarray*}
\fl
\begin{array}{l}
\textrm{Reaction 1: } A \to A+A \quad \textrm{and change the sign of the states},\\
\textrm{Reaction 2: } A + A \to A,
\end{array}
\end{eqnarray*}
i.e., 
\begin{eqnarray*}
\fl
\begin{array}{ll}
n \to n+1 \textrm{ and } \textrm{``$+$ state} \to \textrm{$-$ state''} \,
(\textrm{or} \,\, \textrm{``$-$ state} \to \textrm{$+$ state''}) \quad & \textrm{at rate $\gamma n$},  \\
n \to n-1, \quad 
&\textrm{at rate $\sigma^2 n(n-1)/2$},
\end{array}
\end{eqnarray*}
where $n$ is the number of particles $A$.
Only when the reaction 1 occurs, 
the sign of the state (the internal state) is changed.

Finally, the duality relation for
the above extended stochastic processes is written as follows:
\begin{eqnarray}
\fl
\mathbb{E}_{n,+} \left[  \exp\left\{ \int_0^t V(n_{t'}) \, \rmd t'\right\} z_0^{n_t} \right]
- \mathbb{E}_{n,-} \left[  \exp\left\{ \int_0^t V(n_{t'}) \, \rmd t'\right\} z_0^{n_t} \right]
= \mathbb{E}_{z} \left[ z_t^{n_0} \right],
\label{eq_duality_3}
\end{eqnarray}
where $\mathbb{E}_{n,+}$ and $\mathbb{E}_{n,-}$
mean the expectations with respect to $P(n,+,t)$
and $P(n,-,t)$, respectively.
In addition, as an initial state,
we here set $p(z,0) = \delta(z-z_0)$ for the partial differential equation
and $P(n_0,+,0) = 1$ for the discrete-state stochastic process.
We note that it is possible to extend the duality relations
for arbitrary initial conditions.

\section{Discussions}

In the present paper, the duality concepts were extended for various cases
by using some specific examples.
We here comment on the applicability of the scheme.

Although we restrict ourselves to the stochastic differential equations with only one variable,
multivariate cases can be dealt with adequately.
Actually, in \cite{Ohkubo2010}, the duality relation has been derived based on the Doi-Peliti formalism,
and the multivariate cases have also been discussed;
The introduction of several creation and annihilation operators is enough
to treat the multivariate cases.

When we consider the multivariate cases,
the additional state in section 4 can be reinterpreted in another way.
That is, we introduce a new variable $y$,
which obeys the time-evolution with
\begin{eqnarray}
\frac{\rmd y}{\rmd t} = 0.
\end{eqnarray}
When we set the initial state of $y$ as $y(0) \equiv y_0 = -1$,
\eref{eq_sde_3} can be rewritten as follows:
\begin{eqnarray}
\rmd z = \left( \gamma z + \gamma y z^2 \right) \rmd t + \sigma \sqrt{z (1-z)} \, \rmd W(t),
\label{eq_sde_discussion_1}
\end{eqnarray}
because $y$ is always $-1$.
In the dual process, $y$ is replaced with the creation operator $a^\dagger_y$,
which creates an additional particle corresponding to $y$.
Denoting the dual discrete variables for $z$ and $y$ at time $t$ as
$n^z_t$ and $n^y_t$ respectively,
the duality relation is written as
\begin{eqnarray}
\mathbb{E}_{n^z,n^y} \left[  \exp\left\{ \int_0^t V(n^z_{t'}) \, \rmd t' \right\} z_0^{n^z_t} y_0^{n^y_t} \right]
= \mathbb{E}_{z_0,y_0} \left[ z_t^{n^z_0} y_t^{n^y_0} \right].
\label{eq_duality_discussion}
\end{eqnarray}
Since $y_t = y_0 = -1$,
we can recover \eref{eq_duality_3} by setting $n^y_0 = 0$.

As for the applicability, we must mention about the coefficients of the differential equations.
If the partial differential equation has polynomial (or monomial) coefficients, as discussed in the present paper,
the Doi-Peliti formalism is applicable straightforwardly.
On the other hand, when there are non-polynomial coefficients
we have not yet obtained the range of the applications rigorously.
In order to clarify them, more rigorous mathematical discussions would be needed.
However, for some practical cases, the scheme is applicable, as follows.
For example, we consider the following stochastic differential equation:
\begin{eqnarray}
\rmd s = \sin(s) \rmd t + \rmd W(t).
\label{eq_sde_discussion_2}
\end{eqnarray}
Note that there is the non-polynomial coefficient ($\sin(s)$).
In order to construct the corresponding dual process,
one might consider that the function $\sin(s)$ 
should be replaced with $\sin(a_s^\dagger)$.
However, it is not easy to interpret this term within the dual `stochastic' process.
Instead, we introduce two additional variables:
\begin{eqnarray}
u = \sin(s), \quad v = \cos(s).
\end{eqnarray}
Hence, we have
\begin{eqnarray}
\frac{\rmd s}{\rmd t} = u + \frac{\rmd W(t)}{\rmd t}, \\
\frac{\rmd u}{\rmd t} = \cos(s) \frac{\rmd s}{\rmd t} =  v \frac{\rmd s}{\rmd t}, \\
\frac{\rmd v}{\rmd t} = - \sin(s) \frac{\rmd s}{\rmd t} = - u \frac{\rmd s}{\rmd t},
\end{eqnarray}
and the variables $s$, $u$, and $v$ are interpreted as $a^\dagger_s$, $a^\dagger_u$, and $a^\dagger_v$
in the corresponding dual process.
That is, by using the discrete stochastic process with three variables, $n^s_t$, $n^u_t$, $n^v_t$,
we can construct the dual process for \eref{eq_sde_discussion_2}.

Recently, several works related to the duality have been performed
from physical viewpoint \cite{Borodin2012, Carinci2012}
and mathematical viewpoint \cite{Jansen2012}.
We hope that the extended duality in the present paper motivates further works.
In addition, even when it may be difficult to solve derived discrete-state 
stochastic processes analytically,
the extended duality concepts are useful.
For example,
when one wants to obtain some moments for a stochastic differential equation numerically,
a kind of approximation scheme is needed.
The basic one is the Euler-Maruyama scheme \cite{Kloeden_book},
in which a time-discretization is inevitable.
Hence, in order to make good approximations,
it is necessary to choose an adequate time-discretization.
In contrast, there is no need to use such time-discretization
for simulations of the derived discrete-state stochastic process;
there are various simulation schemes for birth-death processes
without using the time-discretization.
One of the famous algorithms is the Gillespie algorithm \cite{Gillespie1977},
in which time-intervals between events are selected directly from exponential distributions,
so that the time-discretization is not needed.
In addition, the Gillespie algorithm is not an approximate one,
hence rapid and precise calculations could be available
by using the dual stochastic process.

\section*{Acknowledgments}
This work was supported in part by grant-in-aid for scientific research 
(Grant No.~25870339)
from the Ministry of Education, Culture, Sports, Science and Technology (MEXT), Japan.

\end{document}